\documentclass[10pt]{article}
\renewcommand{\theequation}{\thesection.\arabic{equation}}

\begin{document}

\markboth{\protect{\footnotesize\it Noise Transformation in Nonlinear 
System}}{\protect{\footnotesize\it  Noise Transformation in Nonlinear 
System}}

\thispagestyle{empty}

\vspace*{0.37truein}
\baselineskip=13pt
\centerline{\bf NOISE TRANSFORMATION IN NONLINEAR SYSTEM}
\baselineskip=13pt
\centerline{\bf WITH INTENSITY DEPENDENT PHASE ROTATION}
\vspace*{0.37truein}
\centerline{\footnotesize V. V. ZVEREV\footnote{E-mail: zverev@dpt.ustu.ru}}
\baselineskip=12pt
\centerline{\footnotesize \textit{Ural State Technical University, Mira 19, 
620002, Ekaterinburg K-2, Russia}}
\vspace*{0.225truein}

\begin{abstract}
The statistical behavior of a nonlinear system described by a mapping with
phase rotation is studied. We use the Kolmogorov-Chapman equations for the
multi-time probability distribution functions for investigation of dynamics
under the external noise perturbations. We find a stationary solution in the
long-time limit as a power series around a state with complete phase
randomization ("phase mixing"). The Ornstein-Uhlenbeck and Kubo-Andersen
models of noise statistics are considered; the conditions of convergence of
the power expansions are established.

\vspace*{5pt} 
\noindent 2000 \textit{\ Mathematics Subject Classification:}
Primary: 70K55; Secondary: 34F05, 34K50, 37H10,37L55 
\end{abstract}

\section{Introduction}

Noise is a key governing factor in behavior of dissipative nonlinear
systems. Strictly speaking, any purely deterministic description appears to
be inconsistent, because dissipative processes are connected to random
kinetic motion (thermal processes, spontaneous emission, radioactive decay, 
\textit{etc.}), so that they are stochastic in nature. Nevertheless,
deterministic models provide an fair description of averaged motion for
systems with stable trajectories. In the opposite case of unstable motion,
random fluctuations may be amplified, so that validity of non-stochastic
descriptions becomes questionable. The latter case is of particular
importance, because many effects, like noise-induced transitions \cite{b1}
and deterministic diffusion \cite{b2,b3}, are inherent in stochastic
nonlinear systems. It is also worth to mention current attempts to
reconsider the background of statistical mechanics and to demonstrate that
deterministic chaos underlies the origin of macroscopic irreversibility \cite
{b3}-\cite{b6}.

In this work we consider an evolution of a nonlinear noisy discrete
dynamical system. We concentrate on evolutionary mappings with a \textit{%
controllable mixing}, exemplified by the Ikeda mapping \cite{b7}. By this 
is meant that the maximal Lyapunov exponent $L$ is an increasing function 
of a control parameter $\lambda $ and the case $L,\lambda \gg 1$ corresponds 
to chaotic motion with \textit{intense phase mixing} \cite{b8}. The main idea 
of our work is to use $\lambda^{-1}$ as a small parameter and to analyze 
power series around a ''basic'' solution corresponding to the limiting case 
$\lambda \rightarrow \infty $. It is important that in the limit 
$\lambda \rightarrow \infty $ the original nonlinear problem reduces
to a linear one \cite{b8}. Analytical description of such limit transition
is important by itself and it may be useful for validity check of computer
simulations. As it was shown in \cite{b9,b10} for the case of intense phase
mixing, the stationary solutions are associated with certain multifractal
measures.

We present some rigorous results for a nonlinear circuit with delayed
feedback, associated with the Ikeda mapping, which served as a model for
radiation field dynamics in an optical ring cavity with a nonlinear medium 
\cite{b7,b11}. It can be viewed as a particular case of a wide class of
mappings with controllable mixing. For instance, reduction of ordinary
differential equations to mappings (for systems under the action of
piecewise constant or delta-shaped forces) often leads to the mappings with
''nonlinear phase rotations''. In this situation the intense mixing occurs
if a small displacement of a phase point leads to a non small ''rotation''
represented via an element of the orthogonal or unitary group. The 3D
mapping for a nuclear magnetization system \cite{b12,b13}, the 2D mapping
for parametrically driven spin waves in ferromagnetic \cite{b14}, the 2D
mapping for a periodically kicked dissipative oscillator \cite{b15} belong
to this class of mappings.

The random dynamics of the system under consideration is determined by
external noise statistics. Due to the delayed feedback, the noise
contributions from the different times form a superposition. For this reason
it is important to preassign time correlation properties of the external
noise unambiguously.\label{Intro} Here we consider two widely used models of
external noise: the Ornstein-Uhlenbeck \cite{b1} and Kubo-Andersen \cite{b16}
random processes. Both of them may be thought of as an outcome of the
Brownian motion in the thermostat with continuous or discrete assembly of
states.

In the Section 2 we introduce a nonlinear model, write the generalized
Kolmogorov-Chapman \cite{b17} equations for the multi-time distribution
functions and derive the corresponding equations for their Fourier
transforms. In the next Section, we examine the case of intense phase mixing 
$\lambda \rightarrow \infty$ leading to reduced equations; the stationary
solution is found by applying the iterative method. In the Section 4, we find
stationary solutions for the original equations in the form of power series.
In the Section 5, we prove a convergence of series solution and obtain the 
convergence conditions for this series. In the Section 6, we discuss a 
decisive role of the ''rapid'' Gaussian noise.

\section{A basic model}

\label{model}

Consider a circuit with a nonlinear element (NE) and a delayed feedback. We
assume that the transformation of a slowly varying complex valued amplitude $%
X$ consists of the phase shift $\phi \rightarrow \phi +\lambda \left|
X\right| ^2+\Theta _0$, and the dissipative contraction $\left| X\right|
\rightarrow \kappa \left| X\right| $, $\kappa <1$. The signal travels from
the NE through the feedback loop (with a round-trip time $T_d$) and reaches
the mixing device, where it adds to the external signal $1+\xi \left(
t\right) $. Assuming $\xi \left( t\right) $ being a random process with zero
mean value $\left\langle \xi \left( t\right) \right\rangle =0$, we can write
the stochastic difference equation of motion 
\begin{eqnarray}
X\left( t\right) &=&\xi \left( t\right) +F\left( X\left( t-T_d\right) \right)
\label{eq1} \\
&\equiv &\xi \left( t\right) +1+\kappa X\left( t-T_d\right) \exp \left\{
i\lambda \left| X\left( t-T_d\right) \right| ^2+i\Theta _0\right\} . 
\nonumber
\end{eqnarray}
\noindent Equation (\ref{eq1}) may be rewritten in the form of
two-dimensional (complex) mappings family 
\begin{equation}
X_{N+1}=\xi _N+F\left( X_N\right) ,  \label{eq2}
\end{equation}
\noindent where $X_N=X\left( t_0+NT_d\right) $ and $\xi _N=\xi \left(
t_0+\left( N+1\right) T_d\right) $, $0\leq t_0<T_d$. We suppose that the
external noise consist of two components -- one is the Ornstein-Uhlenbeck
random process with small correlation time: $\tau _{cor}\ll T_d$ (hereafter
referred to as the ''rapid'' Gaussian noise); the Fourier transform of its
distribution function $P_G\left( \eta \right) $ is given by $\Lambda \left(
U\right) =\exp \left( -Q\left| U\right| ^2/4\right) $. A second noise
component is assumed to be the Ornstein-Uhlenbeck or Kubo-Andersen Markovian
random process with arbitrary $\tau _{cor}$ and the transition density
functions:

(i) for the Ornstein-Uhlenbeck process

\begin{equation}
W\left( \xi ,\eta ,\left[ \tau \right] \right) =\left\{ \pi R\left( 1-\zeta
_\tau ^2\right) \right\} ^{-1}\exp \left( -\left| \xi -\eta \zeta _\tau
\right| ^2/R\left( 1-\zeta _\tau ^2\right) \right) ,  \label{eq3}
\end{equation}

(ii) for the Kubo-Andersen process

\begin{equation}
W\left( \xi ,\eta ,\left[ \tau \right] \right) =\zeta _\tau \delta ^{\left(
2\right) }\left( \xi -\eta \right) +q\left( \xi \right) \left( 1-\zeta _\tau
\right) ,  \label{eq4}
\end{equation}
\noindent where $\zeta _\tau =\exp \left( -\tau /\tau _{cor}\right) $ , $%
q\left( \xi \right) =\sum\limits_{k=1}^{K_0}p_k\delta ^{\left( 2\right)
}\left( \xi -\xi _k\right) $ and $\delta ^{\left( 2\right) }\left( x\right)
= $ $\delta \left( \mbox{Re}x\right) \delta \left( \mbox{Im}x\right) $ is
the $2D$ (complex) delta-function. The presence of the separate ''rapid''
Gaussian noise component in the external signal is a decisive element in the
convergence proof in Section \ref{convergence}.

Note that the Kubo-Andersen random process (a special case of the kangaroo
process \cite{b16}) is a stepwise constant process describing random jumps
between complex values $\xi _k$, $k=1,2,...,K_0$ (with appearance
probabilities $p_k$); switching times are uniformly and independently
distributed along the time axis.

Under the above-mentioned assumptions we can write the generalized
Kolmogorov-Chapman equation \cite{b8,b17} for the multi-time distribution
functions in the form

\begin{equation}
P_{N+1}\left( \left( X_s\right) ,\zeta \right) =\int d\overline{Y}\;d\xi
\;Q\left( \left( X_s\right) ,\left( Y_s\right) ,\zeta ,\xi \right)
\;P_N\left( \left( Y_s\right) ,\xi \right) ,  \label{eq5}
\end{equation}
\noindent where

\begin{equation}
Q\left( \left( X_s\right) ,\left( Y_s\right) ,\xi _{n+1},\xi _0\right) =\int
d\overline{Y}\;d\bar \eta \;d\xi _1\cdot \cdot \cdot d\xi
_n\prod\limits_{k=0}^nP_G\left( \eta _k\right)  \label{eq6}
\end{equation}

\[
\times \delta ^{\left( 2\right) }\left( X_k-\xi _k-\eta _k-F\left(
Y_k\right) \right) \ \ W\left( \xi _{k+1},\xi _k,\left[ \tau _{k+1}-\tau
_k\right] \right) . 
\]
\noindent In Eq.(\ref{eq5}) we used a short-hand notation

\begin{equation}
P_N\left( \left( X_s\right) ,\xi \right) \equiv P\left( X_0\left[ NT_d+\tau
_0\right] ,X_1\left[ NT_d+\tau _1\right] ,...,X_n\left[ NT_d+\tau _n\right]
,\xi \left[ NT_d\right] \right) ,  \label{eq7}
\end{equation}
\noindent where $X_i$, $i=0,1,...,n$, are the signal amplitudes at $%
NT_d+\tau _i$, respectively, and $\xi $ is the amplitude of the second noise
component at $NT_d$. We also require $0\equiv \tau _0<\tau _1<...<\tau
_{n+1}\equiv T_d$ and use a notation $d\bar A=dA_0\ dA_1...dA_n$.

The corresponding equation for the distribution function Fourier transform

\begin{equation}
\Psi _N\left( \left( U_s\right) ,\Omega \right) =\int d\bar Xd\xi \
P_N\left( \left( X_s\right) ,\xi \right) \exp \left\{ -i\mbox{Re}\left(
\sum\limits_{k=0}^nX_kU_k^{*}+\xi \Omega ^{*}\right) \right\} ,  \label{eq8}
\end{equation}
\noindent is given by

\begin{equation}
\Psi _{N+1}\left( \left( U_s\right) ,\Omega \right) =\int d\bar V\;d\theta \
\Upsilon \left( \left( U_s\right) ,\left( V_s\right) ,\Omega ,\theta \right)
\;\Psi _N\left( \left( V_s\right) ,\theta \right) ,  \label{eq9}
\end{equation}
\noindent where

\begin{equation}
\Upsilon \left( \left( U_s\right) ,\left( V_s\right) ,\Omega ,\theta \right)
=Z\left( \left( U_s\right) ,\Omega ,\theta \right)
\prod\limits_{k=0}^n\Lambda \left( U_k\right) \sigma \left( U_k,V_k\right) ,
\label{eq10}
\end{equation}

\begin{equation}
\sigma \left( U,V\right) =\left( 2\pi \right) ^{-2}\int dY\exp \left\{ i%
\mbox{Re}\left( YV^{*}-F\left( Y\right) U^{*}\right) \right\} ,  \label{eq11}
\end{equation}

\begin{equation}
Z\left( \left( U_s\right) ,\Omega _{n+1},\Omega _0\right) =\int d\Omega
_1...d\Omega _n\prod\limits_{k=0}^nH\left( \Omega _{k+1},\Omega
_k-U_k,\left[ \tau _{k+1}-\tau _k\right] \right) ,  \label{eq12}
\end{equation}

\begin{equation}
H\left( \Omega ,\Gamma ,\left[ \tau \right] \right) =\left( 2\pi \right)
^{-2}\int d\xi d\eta \ W\left( \xi ,\eta ,\left[ \tau \right] \right) \exp
\left\{ i\mbox{Re}\left( \eta \Gamma ^{*}-\xi \Omega ^{*}\right) \right\} .
\label{eq13}
\end{equation}
\noindent Subscripts replacement $N,N+1\rightarrow st$ reduces Eqs.(\ref{eq5}
) and (\ref{eq9}) to the equations for the stationary distributions:

\begin{equation}
P_{st}\left( \left( X_s\right) ,\zeta \right) =\int d\overline{Y}\;d\xi
\;Q\left( \left( X_s\right) ,\left( Y_s\right) ,\zeta ,\xi \right)
\;P_{st}\left( \left( Y_s\right) ,\xi \right) ,  \label{eq14}
\end{equation}

\begin{equation}
\Psi _{st}\left( \left( U_s\right) ,\Omega \right) =\int d\bar V\;d\theta \
\Upsilon \left( \left( U_s\right) ,\left( V_s\right) ,\Omega ,\theta \right)
\;\Psi _{st}\left( \left( V_s\right) ,\theta \right) .  \label{eq15}
\end{equation}

\section{Intense phase mixing case}

\label{limiticase}

The expression (\ref{eq11}) may be presented in the form \cite{b8} 
\begin{equation}
\sigma =\tilde \sigma +\Delta \sigma ,  \label{eq16}
\end{equation}
\noindent where 
\begin{equation}
\sigma \left( U,V\right) =\left( 2\pi \left| V\right| \right) ^{-1}e^{-i%
\mbox{Re}U}\delta ^{\left( 2\right) }\left( \left| V\right| -\kappa \left|
U\right| \right) ,\ \ \Delta \sigma \left( U,V\right) =e^{-i\mbox{Re}U}\beta
\left( U,V\right) ,  \label{eq17}
\end{equation}

\begin{eqnarray}
\beta \left( U,V\right) &=&\frac 1{2\pi \lambda }\sum\limits_{\nu =1}^\infty
\frac 1\nu J_\nu \left( \frac{\kappa \left| UV\right| }{2\lambda \nu }\right)
\label{eq18} \\
&&\ \times \cos \left\{ \nu \left( \Theta _0+\arg \left( \frac VU\right)
\right) -\frac{\left| V\right| ^2+\kappa ^2\left| U\right| ^2}{4\lambda \nu }%
+\frac{\left( \nu +1\right) \pi }2\right\} .  \nonumber
\end{eqnarray}
\noindent In the limit $\lambda \rightarrow \infty $ one can neglect the
last term in Eq.(\ref{eq16}), in other words, replace $\sigma $ by $\tilde
\sigma $ in (\ref{eq10}) (corresponding stationary solutions of Eqs.(\ref
{eq14}), (\ref{eq15}) will be labeled by superscript $^{\left( 0\right) } $%
). Consider the special case of the Ornstein-Uhlenbeck random process in
detail; the Kubo-Andersen process is analized the same way. Using the
explicit form of the function (\ref{eq13}) for the Ornstein-Uhlenbeck
process:

\begin{equation}
H\left( \Omega ,\Gamma ,\left[ \tau \right] \right) =\exp \left\{ -\frac
14\left| \Omega \right| ^2R\left( 1-\zeta _\tau ^2\right) \right\} \delta
^{\left( 2\right) }\left( \Gamma -\Omega \zeta _\tau \right) ,  \label{eq19}
\end{equation}
\noindent one can reduce Eq.(\ref{eq15}) to

\begin{equation}
\Psi _{st}^{\left( 0\right) }=\mathbf{\hat K}\Psi _{st}^{\left( 0\right) }
\label{eq20}
\end{equation}
\noindent where

\[
\mathbf{\hat K}f\left( \left( U_s\right) ,U_{n+1}\right) =\int d\bar
Vd\theta \ f\left( \left( V_s\right) ,\theta \right) Z\left( \left(
U_s\right) ,U_{n+1},\theta \right) \prod\limits_{k=0}^n\Lambda \left(
U_k\right) \tilde \sigma \left( U_k,V_k\right) = 
\]

\begin{equation}
=\prod\limits_{k=0}^n\left[ \Lambda \left( U_k\right) e^{-i\mbox{Re}%
U_k}\right] \Phi \left( \left( U_s\right) ,U_{n+1}\right) \left\langle
f\left( \left( \kappa U_se^{i\phi _s}\right) ,\sum\limits_{j=0}^{n+1}\zeta
_{\tau _j}U_j\right) \right\rangle _{\left( \phi _s\right) },  \label{eq21}
\end{equation}
\noindent and

\begin{equation}
\Phi \left( \left( U_s\right) ,U_{n+1}\right) =\exp \left\{ -\frac
R4\sum\limits_{k=1}^{n+1}\left| \sum\limits_{j=k}^{n+1}\frac{\zeta _{\tau _j}%
}{\zeta _{\tau _k}}U_j\right| ^2\left( 1-\frac{\zeta _{\tau _k}^2}{\zeta
_{\tau _{k-1}}^2}\right) \right\} ,  \label{eq22}
\end{equation}

\[
\zeta _{\tau _k}=e^{-\tau _k/\tau _{cor}},\qquad \left\langle
...\right\rangle _{\left( \phi \right) }\equiv \frac 1{2\pi
}\int\limits_0^{2\pi }\left( ...\right) d\phi . 
\]
\noindent The angular brackets in Eq.(\ref{eq21}) denote an averaging over
the phase angles $\phi _k$, $k=0,1,...,n$. Define a linear operator $\mathbf{%
\hat L}$ acting on the vector $\left[ \left( U_s\right) ,\Omega \right]
\equiv \left[ U_0,...,U_n,\Omega \right] $ as follows 
\begin{equation}
\mathbf{\hat L}\left( \phi _0,...,\phi _n\right) \left[ \left( U_s\right)
,\Omega \right] =\left[ \left( \kappa U_se^{i\phi _s}\right)
,\sum\limits_{k=0}^n\zeta _{\tau _k}U_k+\zeta _{\tau _{n+1}}\Omega \right] .
\label{eq23}
\end{equation}
\noindent It is easy to show that

\begin{equation}
\left\{ \prod\limits_{p=1}^m\mathbf{\hat L}\left( \phi _{0p},...,\phi
_{np}\right) \right\} \left[ \left( U_s\right) ,\Omega \right] =\left[
\left( \kappa ^mU_s\exp \left\{ i\sum_{p=1}^m\phi _{sp}\right\} \right)
,\sum\limits_{k=0}^nd_{km}U_k+\zeta _{\tau _{n+1}}^m\Omega \right] ,
\label{eq24}
\end{equation}
\noindent where

\[
d_{km}=\zeta _{\tau _k}\sum\limits_{q=0}^{m-1}\kappa ^q\zeta _{\tau
_{n+1}}^{m-1-q}\exp \left\{ i\sum\limits_{p=1}^q\phi _{kp}\right\} 
\]
\noindent and $\phi _{kp}$ are arbitrary angles: $0\leq \phi _{kp}<2\pi $, $%
1\leq p\leq m$, $0\leq k\leq n$ (the operators in the product in Eq.(\ref
{eq24}) are enumerated from right to left). One can show that

\begin{equation}
\left| d_{km}\right| \leq \zeta _{\tau _k}\sum\limits_{q=0}^{m-1}\kappa
^q\zeta _{\tau _{n+1}}^{m-1-q}\leq \zeta _{\tau _k}\cdot m\left( \max
\left\{ \kappa ,\zeta _{\tau _{n+1}}\right\} \right) ^{m-1}\stackrel{%
m\rightarrow \infty }{\longrightarrow }0.  \label{eq25}
\end{equation}
\noindent As consequence, we have

\begin{equation}
\lim\limits_{m\rightarrow \infty }\left\{ \prod\limits_{p=1}^m\mathbf{\hat L}%
\left( \phi _{0p},...,\phi _{np}\right) \right\} \left[ \left( U_s\right)
,\Omega \right] =\left[ \left( 0\right) ,0\right]  \label{eq26}
\end{equation}
\noindent and

\begin{equation}
\lim_{m\rightarrow \infty }\mathbf{\hat K}^mf\left( \left( U_s\right)
,\Omega \right) =f\left( \left( 0\right) ,0\right) \cdot \lim_{m\rightarrow
\infty }\mathbf{\hat K}^m\mathbf{1}  \label{eq27}
\end{equation}
\noindent for a continuous function $f$ (in the r.h.s. of Eq.(\ref{eq27} )
the operator acts to the right onto the real constant equal to unity). With
help of the expression (\ref{eq27}), using Eq.(\ref{eq20}) in the iteration
procedure

\begin{equation}
\Psi _{st}^{\left( 0\right) }=\mathbf{\hat K}\Psi _{st}^{\left( 0\right) }=%
\mathbf{\hat K}^2\Psi _{st}^{\left( 0\right) }=\ldots ,  \label{eq28}
\end{equation}
\noindent and taking into account the normalization condition

\begin{equation}
\Psi _{st}^{\left( 0\right) }\left( \left( 0\right) ,0\right) =\int d\bar X\
d\xi \ P_{st}^{\left( 0\right) }\left( \left( X_s\right) ,\xi \right) =1,
\label{eq29}
\end{equation}
\noindent we can write a solution of Eq.(\ref{eq20}) in a simple form

\begin{equation}
\Psi _{st}^{\left( 0\right) }\left( \left( U_s\right) ,\Omega \right)
=\lim_{m\rightarrow \infty }\mathbf{\hat K}^m\mathbf{1}.  \label{eq30}
\end{equation}
\noindent Note that the solution (\ref{eq30}) is the asymptotically stable
''fixed point'' of the evolutionary mapping 
\begin{equation}
\Psi _{N+1}^{\left( 0\right) }=\mathbf{\hat K}\Psi _N^{\left( 0\right) },
\label{eq31}
\end{equation}
\noindent so that it is independent of initial conditions. This reflects
decay of correlations and ''memory loss'' in the system. The transformation $%
P_N\rightarrow P_{N+1}$ defined by Eq.(\ref{eq5}) preserves the norm for
arbitrary $\lambda $. Consequently, the condition $\left| f\right| \leq 1$
implies $\left| \mathbf{\hat K}f\right| \leq 1$, so that absolute values of (%
\ref{eq27}) and (\ref{eq30}) are bounded from above. The operator $\mathbf{%
\hat K}$ performs transformations including \textit{dilation} $\left|
U_s\right| \rightarrow \kappa \left| U_s\right| $ and averaging over the $n-$%
dimensional torus. It is remarkable that we will have Eq. (\ref{eq31}) in
the place of Eq.(\ref{eq9}) if we substitute the \textit{linear} stochastic
difference equation

\begin{equation}
X\left( t\right) =\xi \left( t\right) +1+\kappa X\left( t-T_d\right) \exp
\left\{ i\tilde \theta \left( t\right) \right\}  \label{eq32}
\end{equation}
\noindent instead of the \textit{nonlinear} Eq.(\ref{eq1}); here $\tilde
\theta \left( t\right) $ is the random process with independent and
uniformly distributed values. As it was shown in our work \cite{b10}, the
solutions of a type (\ref{eq30}) are associated with certain integrals over
multifractals.

Using statistical moments approach, one may arrive at the same results in
other way \cite{b8}. Define the stationary moments of distributions as
follows: 
\begin{equation}
M_\gamma ^{st}=\left\langle \xi ^p\xi ^{*q}\prod\limits_{k=0}^n\left(
X_k^{r_k}X_k^{*l_k}\right) \right\rangle _{st}=D_\gamma \left( \Psi
_{st}^{\left( 0\right) }\right) ,  \label{eq33}
\end{equation}
\noindent where multiple derivatives

\begin{equation}
D_\gamma \left( f\right) \stackrel{def}{=}\frac{\partial ^p}{\partial \Omega
^{*p}}\frac{\partial ^q}{\partial \Omega ^q}\prod\limits_{k=0}^n\left.
\left( \frac{\partial ^{r_k}}{\partial U_k^{*r_k}}\frac{\partial ^{l_k}}{%
\partial U_k^{l_k}}\right) f\left( \left( U_s\right) ,\Omega \right) \right|
_{{\left( U_s\right) =\mathbf{0},\Omega =0}}  \label{eq34}
\end{equation}
\noindent are computed at the origin, and $\gamma =\left( p,q,\left(
r_s,l_s\right) \right) $ denotes the multi-index. It follows from Eq. (\ref
{eq21}) that $D_\gamma \left( \mathbf{\hat K}f\right) $ is a linear
combination of $D_{\gamma ^{^{\prime }}}\left( f\right) $ for $\left| \gamma
^{^{\prime }}\right| <\left| \gamma \right| $ (here we denote $\left| \gamma
\right| =p+q+\sum\limits_{k=0}^n\left( r_k+l_k\right) $). Introduce a vector 
$\mathbf{m}\left( f\right) $ of values $D_\gamma \left( f\right) $, $\left|
\gamma \right| <\left| \gamma _{max}\right| $, and choose special ordering
of its components: provided that either $\left| \gamma _2\right| <\left|
\gamma _1\right| $, or $\left| \gamma _2\right| =\left| \gamma _1\right| $
with $p_2+q_2>p_1+q_1$, \textbf{$D_{\gamma _1}\left( f\right) $ }precedes%
\textbf{\ $D_{\gamma _2}\left( f\right) $: } 
\begin{equation}
\mathbf{m}\left( f\right) =\left( ...,D_{\gamma _1}\left( f\right)
,...,D_{\gamma _2}\left( f\right) ,...\right) .  \label{eq35}
\end{equation}
\noindent It is easy to see that 
\begin{equation}
\mathbf{m}\left( \mathbf{\hat K}f\right) =\mathbf{am}\left( f\right) +%
\mathbf{b},\qquad \mathbf{b}=\mathbf{b}_0\cdot f\left( \left( 0\right)
,0\right) ,  \label{eq36}
\end{equation}
\noindent where $\mathbf{a}$ is a\textbf{\ }triangular matrix. The diagonal
elements

\begin{equation}
\mathbf{a}_{\gamma \gamma }=\Xi _{pq}\prod\limits_{k=0}^n\delta
_{r_{k\;}l_k}\kappa ^{r_k+l_k},  \label{eq37}
\end{equation}
with $\Xi _{pq}=e^{-T_d/\tau _{cor}}+\left( 1-e^{-T_d/\tau _{cor}}\right)
\delta _{p0}\delta _{q0}$ for the Kubo-Andersen process and $\Xi
_{pq}=e^{-T_d\left( p+q\right) /\tau _{cor}}$ for the Ornstein-Uhlenbeck
process, are real and $0<\mathbf{a}_{\gamma \gamma }<1$, so that both $%
\mathbf{a}$ and $\mathbf{1}-\mathbf{a}$ are nonsingular matrices.
Furthermore, $\mathbf{a}=\mathbf{TJT}^{-1}$, where $\det \mathbf{T}\neq 0$
and $\mathbf{J}$ is the Jordan form of the matrix $\mathbf{a}$ (each
diagonal element of $\mathbf{J}$ also is positive and less than unity). In
this case $\mathbf{a}^m=\mathbf{TJ}^m\mathbf{T}^{-1}\stackrel{m\rightarrow
\infty }{\longrightarrow }0$ and

\[
\lim_{m\rightarrow \infty }\left\{ \mathbf{m}\left( \mathbf{\hat K}%
^mf\right) -\left( \mathbf{1}-\mathbf{a}\right) ^{-1}\mathbf{b}_0\cdot
f\left( \left( 0\right) ,0\right) \right\} 
\]

\begin{equation}
=\lim_{m\rightarrow \infty }\mathbf{a}^m\left\{ \mathbf{m}\left( f\right)
-\left( \mathbf{1}-\mathbf{a}\right) ^{-1}\mathbf{b}_0\cdot f\left( \left(
0\right) ,0\right) \right\} =0.  \label{eq38}
\end{equation}
\noindent Matrix relations (\ref{eq36}) for various $\gamma _{max}$ present
another equivalent form of the condition (\ref{eq27}).

\section{Series expansion of exact solution}

\label{expansion}

In the general case of arbitrary values of $\lambda $ we can rewrite Eq.(\ref
{eq15}) as 
\begin{equation}
\Psi _{st}=\left( \mathbf{\hat K}+\varepsilon \mathbf{\hat S}\right) \Psi
_{st},  \label{eq39}
\end{equation}
\noindent where $\varepsilon $ is an auxiliary parameter. We are seeking a
solution of Eq.(\ref{eq39}) in the form of formal power series in $%
\varepsilon $ \cite{b18}: 
\begin{equation}
\Psi _{st}=\sum\limits_{j=0}^\infty \varepsilon ^j\Psi _{st}^{\left(
j\right) }  \label{eq40}
\end{equation}
\noindent with normalization conditions 
\begin{equation}
\Psi _{st}\left( \left( 0\right) ,0\right) =\Psi _{st}^{\left( 0\right)
}\left( \left( 0\right) ,0\right) =1,  \label{eq41}
\end{equation}
\begin{equation}
\Psi _{st}^{\left( j\right) }\left( \left( 0\right) ,0\right) =0,\qquad j>0.
\label{eq42}
\end{equation}
\noindent Substituting the expansion (\ref{eq40}) into Eq.(\ref{eq39}) and
equating coefficients of like powers of $\varepsilon $, we obtain Eq.(\ref
{eq20}) and

\begin{equation}
\Psi _{st}^{\left( j\right) }=\mathbf{\hat K}\Psi _{st}^{\left( j\right) }+%
\mathbf{\hat S}\Psi _{st}^{\left( j-1\right) }.  \label{eq43}
\end{equation}
\noindent Repeated application of the last relation yields 
\begin{equation}
\Psi _{st}^{\left( j\right) }=\mathbf{\hat K}^m\Psi _{st}^{\left( j\right)
}+\sum\limits_{p=0}^{m-1}\mathbf{\hat K}^p\mathbf{\hat S}\Psi _{st}^{\left(
j-1\right) }  \label{eq44}
\end{equation}
\noindent where $m$ is an arbitrary positive integer. In the limit $%
m\rightarrow \infty $ in Eq.(\ref{eq44}), using Eqs.(\ref{eq27}), (\ref{eq42}%
) and its corollary: 
\begin{equation}
\lim_{m\rightarrow \infty }\mathbf{\hat K}^m\Psi _{st}^{\left( j\right)
}=0,\qquad j>0,  \label{eq45}
\end{equation}
\noindent one can find a recursive relation 
\begin{equation}
\Psi _{st}^{\left( j\right) }=\sum\limits_{p=0}^\infty \mathbf{\hat K}^p%
\mathbf{\hat S}\Psi _{st}^{\left( j-1\right) }.  \label{eq46}
\end{equation}
\noindent Finally, combination of recursions (\ref{eq46}) with different $j$
and substitution into Eq.(\ref{eq40}) yields 
\begin{equation}
\Psi _{st}=\Psi _{st}^{\left( 0\right) }+\sum\limits_{j=1}^\infty
\varepsilon ^j\left\{ \left( \sum\limits_{p=0}^\infty \mathbf{\hat K}%
^p\right) \mathbf{\hat S}\right\} ^j\Psi _{st}^{\left( 0\right) }.
\label{eq47}
\end{equation}
\noindent In the case $\lambda \gg 1$ the term with $\varepsilon ^j$ in Eq.(%
\ref{eq47}) is $O\left( \lambda ^{-2j}\right) $; note that one must set $%
\varepsilon =1$ in the final expression.

\section{A proof of convergence}

\label{convergence}

In this section we find a condition of existence of the expansion (\ref{eq47}%
). Consider the associated equation for the ''fixed point'' $\Psi _{st}$:

\begin{equation}
\Psi _{st}=\mathbf{\hat A}\Psi _{st},  \label{eq48}
\end{equation}
for the operator: 
\begin{equation}
\mathbf{\hat A}f=\Psi _{st}^{\left( 0\right) }+\left\{
\sum\limits_{p=0}^\infty \mathbf{\hat K}^p\right\} \mathbf{\hat S}f.
\label{eq49}
\end{equation}
\noindent The operator $\mathbf{\hat K}$ is defined by Eq.(\ref{eq21}) and

\begin{equation}
\mathbf{\hat S}f\left( \left( U_s\right) ,\Omega \right)
=\prod\limits_{k=0}^n\Lambda \left( U_k\right) \int d\bar Vd\theta \ f\left(
\left( V_s\right) ,\theta \right) Z\left( \left( U_s\right) ,\Omega ,\theta
\right) \alpha \left( \left( U_s\right) ,\left( V_s\right) \right) ,
\label{eq50}
\end{equation}
\noindent where 
\begin{equation}
\alpha \left( \left( U_s\right) ,\left( V_s\right) \right)
=\sum\limits_{k=0}^n\frac 1{\left( k+1\right) !\left( n-k\right)
!}\sum\limits_{\left( r_s\right) }\left[ \prod\limits_{i=0}^k\Delta \sigma
\left( U_{r_i},V_{r_i}\right) \prod\limits_{j=k+1}^n\tilde \sigma \left(
U_{r_j},V_{r_j}\right) \right] .  \label{eq51}
\end{equation}
\noindent The inner summation in (\ref{eq51}) is taken over all the
permutations 
\[
\left( 
\begin{array}{ccccc}
0 & 1 & 2 & ... & n \\ 
r_0 & r_1 & r_2 & ... & r_n
\end{array}
\right) . 
\]
\noindent For $k=n$ the right-hand product of factors $\tilde \sigma $ in (%
\ref{eq51}) is equal to unity. We are going to prove that $\mathbf{\hat A}$
is the \textit{contraction operator} in an appropriate \textit{complete
metric space.} The condition of contraction is given by 
\begin{equation}
\left\| \mathbf{\hat A}f_2-\mathbf{\hat A}f_1\right\| \leq \ \alpha
_{con}\left\| f_2-f_1\right\| ,\qquad \alpha _{con}<1.  \label{eq52}
\end{equation}
\noindent According to the \textit{contraction mapping principle}, Eq.(\ref
{eq48}) has the \textit{unique solution} if the condition (\ref{eq52}) is
satisfied. From Eq.(\ref{eq49}) follows

\begin{equation}
\mathbf{\hat A}f_2-\mathbf{\hat A}f_1=\left\{ \sum\limits_{p=0}^\infty 
\mathbf{\hat K}^p\right\} \mathbf{\hat S(}f_2-f_1)  \label{eq53}
\end{equation}
\noindent and the operator in the right hand side of (\ref{eq53}) is linear.
In this case the condition (\ref{eq52}) takes place simultaneously with the
limiting condition for the operator norm

\begin{equation}
\left\| \left\{ \sum\limits_{p=0}^\infty \mathbf{\hat K}^p\right\} \mathbf{%
\hat S}\right\| \stackrel{def}{=}\sup_{f\neq 0}\left( \frac 1{\left\|
f\right\| }\left\| \left\{ \sum\limits_{p=0}^\infty \mathbf{\hat K}%
^p\right\} \mathbf{\hat S}f\right\| \right) <\alpha _{con}<1.  \label{eq54}
\end{equation}
\noindent The expansion (\ref{eq47}) is convergent under the condition (\ref
{eq54}). On the other hand, (\ref{eq47}) is the solution of Eq.(\ref{eq39})
resulting from the iterative procedure.

In this work we use the norm 
\begin{equation}
\left\| f\right\| \stackrel{def}{=}\sup_{\left( \left( U_s\right) ,\Omega
\right) \in \mathbf{C}^{n+2}}\left\{ \left| f\left( \left( U_s\right)
,\Omega \right) \right| \prod\limits_{k=0}^n\mu \left( U_k\right) \right\}
\label{eq55}
\end{equation}
\noindent  with 
\[
\mu \left( U\right) =\left\{ \Lambda \left( U\right) \right\} ^{-1/2}=\exp
\left\{ \frac 18Q\left| U\right| ^2\right\} . 
\]
\noindent  (properties of this norm are discussed in the Appendix). To prove
the inequalities (\ref{eq52}),(\ref{eq54}) and make estimate of $\alpha
_{con}$, we have to make a number of auxiliary estimations. Using Eq.(\ref
{eq19}) and the similar expression for the Kubo-Andersen process: 
\begin{equation}
H\left( \Omega ,\Gamma ,\left[ \tau \right] \right) =\zeta _\tau \delta
^{\left( 2\right) }\left( \Omega -\Gamma \right) +\left( 1-\zeta _\tau
\right) \chi \left( \Omega \right) \delta ^{\left( 2\right) }\left( \Gamma
\right) ,  \label{eq56}
\end{equation}
\noindent where 
\[
\chi \left( \Omega \right) =\sum\limits_{k=1}^{K_0}p_k\exp \left\{ -i%
\mbox{Re}\left( \xi _k\Omega ^{*}\right) \right\} 
\]
\noindent (see Eq.(\ref{eq4})), one can find for both noise statistics

\begin{equation}
\sup_{\Omega \in \mathbf{C}}\left| \int d\theta \ H\left( \Omega ,\theta
-U,\left[ \tau \right] \right) \ f\left( U,\theta \right) \right| \leq
\sup_{\Omega \in \mathbf{C}}\left| f\left( U,\Omega \right) \right| ,
\label{eq57}
\end{equation}
\begin{equation}
\sup_{\Omega \in \mathbf{C}}\left| \int d\theta \ Z\left( \left( U_s\right)
,\Omega ,\theta \right) \ f\left( \left( U_s\right) ,\theta \right) \right|
\leq \sup_{\Omega \in \mathbf{C}}\left| f\left( \left( U_s\right) ,\Omega
\right) \right| ;  \label{eq58}
\end{equation}
\noindent $f\left( U,\Omega \right) $ in (\ref{eq57}) and $f\left( \left(
U_s\right) ,\Omega \right) $ in (\ref{eq58}) are arbitrary bounded complex
functions. Next, using (\ref{eq21}), we can obtain 
\begin{equation}
\sup_{\Omega \in \mathbf{C}}\left| \mathbf{\hat K}^jf\left( \left(
U_s\right) ,\Omega \right) \right| \leq
\prod\limits_{k=0}^n\prod\limits_{m=0}^{j-1}\Lambda \left( \kappa
^mU_k\right) \left\langle \sup_{\Omega \in \mathbf{C}}\left| f\left( \left(
\kappa ^jU_se^{i\phi _s}\right) ,\Omega \right) \right| \right\rangle
_{\left( \phi _s\right) }.  \label{eq59}
\end{equation}
\noindent Besides, using (\ref{eq18}) and the inequalities $\left| J_\nu
\left( z\right) \right| \leq \frac 1{\nu !}\left| z/2\right| ^\nu $ for real
values of $z$ and integer positive $\nu $ \cite{b19}; $\kappa ^{\nu \left(
j+1\right) }\leq \kappa ^{j+1}$, $\nu ^{-\nu -1}\leq 1$ for $\nu \geq 1$, $%
j\geq 0$, $\kappa <1$, we can find the upper bounds 
\begin{equation}
\left\langle \left| \beta \left( \kappa ^jUe^{i\phi },V\right) \right|
\right\rangle _{\left( \phi \right) }\leq \frac{\kappa ^{j+1}}{2\pi \lambda }%
\exp \left\{ \frac{\left| UV\right| }{4\lambda }\right\} \leq \frac{\kappa
^{j+1}}{2\pi \lambda }\exp \left\{ \frac{\left| U\right| ^2+\left| V\right|
^2}{8\lambda }\right\} .  \label{eq60}
\end{equation}
\noindent Finally, the inequalities (\ref{eq59}), (\ref{eq60}) are combined
to produce the \textit{contraction condition} (\ref{eq52}) with 
\begin{equation}
\alpha _{con}=\frac 1{1-\kappa }\left\{ \left( \frac{4\kappa }{Q\lambda -1}%
+1\right) ^{n+1}-1\right\} .  \label{eq61}
\end{equation}
\noindent The parameter $\alpha _{con}$ is less than unity for any chosen $n$
if $Q\lambda $ is sufficiently large. Under this condition, the solution 
(\ref{eq47}) of Eq. (\ref{eq15}) \textit{exists} as the \textit{unique} 
''fixed point'' of the contraction operator (\ref{eq49}). It is possible to 
prove convergence of (\ref{eq47}) in another way, using the technique of a 
majorizing series, but it's very cumbersome and we omit it here.

\section{Conclusions and discussion}

We have found a stationary solution to the generalized Kolmogorov-Chapman
equation for the multi-time distribution functions in the form of a 
convergent series expansion (\ref{eq47}). Our approach is based on
successive application of two iterative procedures. The final expressions
were obtained for the case of the nonlinear Ikeda mapping and certain 
fluctuation statistics. This result may be used for calculation of the 
approximate expressions for probability distributions, statistical moments 
and correlation functions.

As simple example consider a case with the ''rapid'' Gaussian noise
component only. In this case Eq.(\ref{eq15}) for the Fourier transform of
the one-time distribution function takes the form: 
\begin{equation}
\Psi _{st}\left( U\right) =\Lambda \left( U\right) \int dV\ \Psi _{st}\left(
V\right) \sigma \left( U,V\right)  \label{eq62}
\end{equation}
Using the general expression (\ref{eq47}), we can write approximate
first-order solution as follows: 
\[
\Psi _{st}\left( U\right) \cong \Psi _{st}^{\left( 0\right) }\left( U\right)
+\left\{ \sum\limits_{p=0}^\infty \mathbf{\hat K}^p\right\} \mathbf{\hat S}%
\Psi _{st}^{\left( 0\right) }\left( U\right) =e^{-i\mbox{Re}U}\Lambda \left(
U\right) \{\eta _\infty \left( \kappa \left| U\right| \right) 
\]
\begin{equation}
+\int\limits_0^\infty vdv\Lambda \left( v\right) \eta _\infty \left( \kappa
v\right) \left( \beta _1\left( U,v\right) +\sum\limits_{p=1}^\infty \Lambda
\left( \kappa ^pU\right) \eta _{p-1}\left( \kappa \left| U\right| \right)
\beta _2\left( \kappa ^p\left| U\right| ,v\right) \right) \},  \label{eq63}
\end{equation}
\noindent where 
\begin{equation}
\beta _1\left( U,v\right) =2\pi \left\langle \beta \left( U,ve^{i\phi
}\right) \exp \left\{ -iv\cos \phi \right\} \right\rangle _{\left( \phi
\right) },  \label{eq64}
\end{equation}
\begin{equation}
\beta _2\left( u,v\right) =2\pi \left\langle \beta \left( ue^{i\phi
_1},ve^{i\phi _2}\right) \exp \left\{ -i\left( u\cos \phi _1+v\cos \phi
_2\right) \right\} \right\rangle _{\left( \phi _1,\phi _2\right) },
\label{eq65}
\end{equation}
\begin{equation}
\eta _k\left( u\right) =\prod\limits_{j=0}^{k-1}\Lambda \left( \kappa
^ju\right) \ J_0\left( \kappa ^ju\right) ,\qquad \eta _0=1,\qquad \eta
_\infty =\lim_{k\rightarrow \infty }\eta _k.  \label{eq66}
\end{equation}
\noindent In general case the resulting formulae are cumbersome.
Nevertheless, they provide a basis for development of computational
procedures (in particular, using symbolic processors). The detailed study of
this applications is beyond the scope of this article.

Consider a specific role of the ''rapid'' Gaussian noise in our method. In
order to simplify consideration, we suppose that this is the only noise
component present in the system. To analize influence of noise, we replace $%
\sigma \left( U,V\right) $ by $\Lambda \left( U\right) \sigma \left(
U,V\right) $ in Eq.(\ref{eq62}). Since $\Lambda \left( U\right) =\exp \left(
-\frac 14Q\left| U\right| ^4\right) $, the Fourier spectrum becomes
truncated in the region $\left| U\right| \gg 1$. It ensures the convergence
of the expansion (\ref{eq47}). On the other hand, as it was shown in \cite
{b8}, this replacement leads to a ''coarse-grained'' dynamics, because
fluctuations coarsen the fine-scale structure of a chaotic attractor. If $%
Q\lambda $ is sufficiently large, the condition of contraction (\ref{eq52})
may be satisfied for any $n$, so that corresponding expansions (\ref{eq40}),
(\ref{eq47}) converge. However, we can consider the limits $\lambda
\rightarrow \infty $ and $Q\rightarrow 0$ separately. The limits do not
commute, so that we should set $Q\rightarrow 0$ after using $\lambda
\rightarrow \infty $. It seems plausible that the correct description of the
random behavior in the limit $\lambda \rightarrow \infty $ may be reached
only under the action of an infinitesimal noise.

It is instructive to draw a parallel between the above manipulations and the
well-known formalism of quantum mechanics. In order to obtain the retarded
(causal) solutions of the Schrodinger equation, one have to write the
equation containing the additional source term: 
\begin{equation}
\ddot \psi _\epsilon +i\left( \mathbf{\hat H}_0+\mathbf{\hat V}\right) \psi
_\epsilon =-\epsilon \left( \psi _\epsilon -\Phi \right) ,  \label{eq67}
\end{equation}
\noindent where $\Phi $ is a solution for the ''nonperturbed'' system $%
\left( \mathbf{\hat V}=0\right) $, and further to set $\epsilon \rightarrow
0 $ after the volume grows infinitely (this is the boundary condition of
Gell-Mann and Goldberger \cite{b20} in the form proposed in \cite{b21}; here 
$\hbar =0$). The same approach appears to be natural to formulate one
well-known version of the so-called ''master equation'' formalism in
statistical physics. Following the technique of \cite{b21}, the final
equation of motion for the quasi-equilibrium statistical operator may be
derived from the Liouwille equation (completed with the source term) by
substituting $\epsilon \rightarrow 0$ after the thermodynamic limit is
reached. Thus, in order to obtain a physically significant irreversible
solution, one must add the infinitesimal dissipative term (in other words,
to break the symmetry with respect to $t\rightarrow -t$).

Both above-mentioned procedures lead to a ''coarsened'' description,
presented by ''coarsened'' equations. It is worth to note that dissipative
processes accompanied by fluctuations bring into existence a strange
attractor the same time destroying a fine structure of the attractor at $%
\lambda \rightarrow \infty $ (for an example of the attractor in our model,
see \cite{b22}) .

\section*{Appendix}

\renewcommand{\theequation}{A.\arabic{equation}}

In this work we use a nonstandard metrics, so it is necessary to check the
completeness of the respective metric space. Define a norm and a distance
function on a set of functions of $n$ complex arguments $\mathbf{x}=\left(
x_1,x_2,...,x_n\right) \in \mathbf{C}^n$ as follows: 
\begin{equation}
\left\| f\right\| _\infty =\sup_{\mathbf{x}\in \mathbf{C}^n}\left| f\left( 
\mathbf{x}\right) \right| ,\qquad \rho _\infty \left( f_1,f_2\right)
=\left\| f_1-f_2\right\| _\infty .  \label{eq68}
\end{equation}
\noindent Using the standard approach (see, for example, \cite{b23}, Theorem
I.23), we can prove

(i) if $f_k\left( \mathbf{x}\right) $, $k\in \mathbf{N}$, is a Cauchy
sequence, the pointwise and uniform convergence takes place:

\begin{equation}
\left( \forall \mathbf{x}\in \mathbf{C}^n\right) \left( \forall \varepsilon
>0\right) \left( \exists N_\varepsilon \right) \left( \forall
k>N_\varepsilon \right) :\left| f_k\left( \mathbf{x}\right) -f\left( \mathbf{%
x}\right) \right| <\varepsilon ;  \label{eq69}
\end{equation}

(ii) if $f_k\left( \mathbf{x}\right) $ is continuous for each $k\in \mathbf{N%
}$, $f\left( \mathbf{x}\right) $ is continuous as well.

\noindent In addition, we can prove:

(iii) if all $f_k\left( \mathbf{x}\right) $ are uniformly bounded

\begin{equation}
\left( \exists M\right) \left( \forall k\in \mathbf{N}\right) \left( \forall 
\mathbf{x}\in \mathbf{C}^n\right) :\left| f_k\left( \mathbf{x}\right)
\right| \leq M,  \label{eq70}
\end{equation}
\noindent then $\left| f\left( \mathbf{x}\right) \right| \leq M$.

It is convenient to prove (iii) by reduction to contradiction. Assume that

\begin{equation}
\left( \exists \mathbf{x}_0\in \mathbf{C}^n\right) :\left| f\left( \mathbf{x}%
_0\right) \right| =M+\bar \varepsilon ,\qquad \bar \varepsilon >0.
\label{eq71}
\end{equation}
\noindent Choosing $\varepsilon <\bar \varepsilon $ and $\bar k$
sufficiently large to have $\left| f_{\bar k}\left( \mathbf{x}_0\right)
-f\left( x_0\right) \right| <\varepsilon $, we arrive at a contradiction to
the original assumption (\ref{eq71}): 
\begin{equation}
\left| f_{\bar k}\left( \mathbf{x}_0\right) \right| \geq \left| \left|
f\left( \mathbf{x}_0\right) \right| -\left| f_{\bar k}\left( \mathbf{x}%
_0\right) -f\left( \mathbf{x}_0\right) \right| \right| \geq M+\bar
\varepsilon -\varepsilon >M.  \label{eq72}
\end{equation}
\noindent It means that the continuous uniformly bounded functions, equipped
with the norm and metric (\ref{eq68}), form the complete metric space.

It follows that the space $C^{\star }\subset C\left( \mathbf{C}^{n+2}\right) 
$ of continuous functions, satisfying 
\begin{equation}
\left| f\left( \left( U_s\right) ,\Omega \right) \right| \leq
M\prod\limits_{k=0}^n\mu ^{-1}\left( U_k\right) ,  \label{eq73}
\end{equation}
\noindent with the norm (\ref{eq55}) and the distance $\rho _{C\star }\left(
f_1,f_2\right) =\left\| f_1-f_2\right\| $, is the \textit{complete metric
space} (the condition (\ref{eq73}) is rewritten as $\left\| f\right\| \leq M$%
).

Finally, note that $f\in C^{\star }$ implies $Af\in C^{\star }$. Choosing a
class of functions with $\left\| f\right\| \leq M$, where $M$ is
sufficiently large: $M>\left( 1-\alpha _{con}\right) ^{-1}$ (the functional
space $C^{\star }$ may be expanded by replacement $M\rightarrow M^{^{\prime
}}$, $M^{^{\prime }}>M$, if required), we may find that 
\begin{equation}
\left\| Af\right\| \leq \left\| \Psi _{st}^{\left( 0\right) }\right\|
+\left\| \left( \sum\limits_{p=0}^\infty \mathbf{\hat K}^p\right) \mathbf{%
\hat S}f\right\| \leq 1+\alpha _{con}\left\| f\right\| \leq 1+\alpha
_{con}M<M.  \label{eq74}
\end{equation}

\section*{Acknowledgments}

The author wants to thank B.Y. Rubinstein for help and discussions. The work
was supported in part by grant of RF State Committee on High Education
95-0-8.3-14.

\end{document}